\documentclass[a4paper,11pt]{article}
\usepackage[utf8]{inputenc}
\usepackage{color}
\usepackage{cite}
\usepackage{hyperref}
\usepackage{amsmath}
\usepackage{graphicx}

\usepackage[top=0.4in, bottom=0.7in, left=0.60in, right=0.5in]{geometry}

\begin{document}
\title{{\Large Exclusive production of P-wave charmonia through two virtual photons in $e^-+e^+$ annihilation}}
\author{Avinash Okram$^1$, Shashank Bhatnagar$^1$}
\maketitle \small{1. Department of Physics, University Institute of Sciences, Chandigarh University, Mohali-140413, India\\}

\begin{abstract}
\normalsize{We present a relativistic study of the exclusive double–charmonium processes $e^- e^+ \rightarrow \gamma^* \gamma^* \rightarrow \chi_{c1}+\eta_c$, and $e^- e^+ \rightarrow \gamma^* \gamma^* \rightarrow h_c+ h_c$ at $\sqrt{s}=10.6$ GeV in the framework of $4\times 4$ Bethe-Salpeter equation. Since these channels are forbidden in single-photon annihilation, they proceed purely through two-photon quark-rearrangement mechanisms and thus provide a clean probe of relativistic quarkonium dynamics. Using parameters fixed from mass spectroscopy from our earlier works, we obtain leading-order predictions for total cross sections and their energy dependence. The predicted cross sections are of order $10^{-3}$fb and exhibit a hierarchy that originates from the radial structure of the meson wave functions, which determines the overlap integrals contributing to the production amplitude. While the absolute magnitudes of cross sections differ from NRQCD estimates, the qualitative ordering of channels is consistent with NRQCD expectations. We find a smooth power-law falloff of the cross section for both $\chi_{c1}+\eta_c$ and $h_c+h_c$. These results provide benchmark predictions for future searches at high-luminosity $e^-+e^+$ colliders.}
\end{abstract}

\section{Introduction}
Exclusive double-charmonium production in $e^- e^+$ annihilation at B-factories remains a lively testing ground for our understanding of heavy-quark dynamics, relativistic bound-state effects, and the interplay of QED and QCD mechanisms at intermediate energies. Early Belle and BABAR measurements of channels such as $e^- e^+\rightarrow \gamma^*\rightarrow J/\Psi+\eta_c$ and $e^- e^+\rightarrow \gamma^*\rightarrow J/\Psi+\chi_{cJ}$ \cite{belle02,belle04,babar05} showed large discrepancies with leading-order NRQCD predictions\cite{braaten03,bodwin95}. Subsequent studies demonstrated that a consistent description requires the simultaneous inclusion of higher-order QCD corrections, relativistic effects, and contributions beyond the simplest nonrelativistic approximations \cite{braaten03a,bodwin06,wang08,huang23,fan12,braaten}.  This included processes with final states forbidden through a single virtual photon, but allowed through two-photon exchange, such as $e^- e^+\rightarrow \gamma^*\gamma^*\rightarrow J/\Psi+J/\Psi$, and $e^- e^+\rightarrow \gamma^*\gamma^*\rightarrow \eta_c+\eta_c$, which proceed at order in $\alpha^4$. This is allowed only for final states with identical charge conjugation parity. Thus $J/\Psi+J/\Psi$ \cite{braaten03a,luchinsky,peskin}and $\eta_c+\eta_c$\cite{braaten03a} production through two virtual photons is suppressed by additional powers of $\alpha_{em}$ relative to single–photon channels. This suppression of $\eta_c+\eta_c$ cross section through two intermediate photons was also noticed in our recent calculation of this process in framework of Bethe-Salpeter equation\cite{bhatnagar25}.

However enhancements from photon-fragmentation kinematics (in case of $J/\Psi+J/\Psi$), collinear logarithms, and quark-rearrangement diagrams can render their cross sections phenomenologically relevant at B-factory energies. This competition between suppression due to QED coupling, and kinematic enhancement has been extensively studied within NRQCD\cite{braaten03}, light-cone approaches\cite{braguta07}, vector-dominance model\cite{fan12}, and other frameworks. In our recent work, we investigated the above two processes\cite{bhatnagar25} at $\sqrt{s}$=10.6 GeV within the fully relativistic $4\times 4$ Bethe–Salpeter equation (BSE) framework under the covariant instantaneous ansatz. We had also previously evaluated the production processes, $e^- e^+\rightarrow\gamma^*\rightarrow \gamma+\chi_{cJ}$, and $e^+\rightarrow \gamma^*\rightarrow\gamma+\eta_c$\cite{bhatnagar24}; $e^- e^+\rightarrow\gamma^*\rightarrow h_c+\chi_{c1}$, and $e^- e^+\rightarrow\gamma^*\rightarrow h_c+\eta_c$\cite{monika23}, as well as $e^-e^+\rightarrow \gamma^*\rightarrow \gamma+J/\Psi$\cite{bhatnagar25a}. These studies in the $4\times 4$ Bethe–Salpeter framework under the Covariant Instantaneous Ansatz (CIA) have shown that BSE’s relativistic vertex structure and its consistent linkage between spectroscopy and transition amplitudes (without introducing additional parameters) can produce sizable leading-order contributions for several exclusive production channels at $\sqrt{s}$=10.6 GeV that compare favourably with experimental upper limits where available.

In the present work, we extend our systematic investigation of double charmonium production through two virtual photons to the processes: $e^- e^+\rightarrow\gamma^*\gamma^*\rightarrow \chi_{c1}+\eta_c$, and $e^- e^+\rightarrow\gamma^*\gamma^*\rightarrow h_c+h_c$ at $\sqrt{s}$=10.6 GeV within the same $4\times 4$ BSE framework. The process  $e^- e^+\rightarrow\gamma^*\gamma^*\rightarrow \chi_{c1}+\eta_c$ involves production of two charmonia with opposite spin configurations but identical charge parity ($C=+1$), while $e^-e^+\rightarrow\gamma^*\gamma^*\rightarrow h_c+h_c$ corresponds to the production of two axial–vector spin-singlet P-wave states. Thus, these channels probe C-even charmonium production through $\gamma^*\gamma^*$ complementary to: single-photon NRQCD-dominated processes, and gluon-rich hadroproduction. $h_ch_c$ production is a clean test of spin-singlet P-wave dynamics that is rarely accessible experimentally, while $\chi_{c1}\eta_c$ test axial–pseudoscalar correlations that are absent in $J/\Psi J/\Psi$ productions. Although these channels have not yet been observed, future Belle II searches can set stringent upper limits on their production rates. Such limits would provide valuable constraints on relativistic quark models and Bethe Salpeter equation–based descriptions of P-wave charmonium production.

Now the photon fragmentation mechanism is forbidden in these channels by C-parity and spin–parity constraints, leading to qualitatively different dynamical behavior and suppression of the kinematical enhancements characteristic of vector–vector final states. Both reactions thus provide a clean probe of quark rearrangement dynamics in two-photon processes that are free from fragmentation contributions (as in case of $J/\Psi+J/\Psi$\cite{braaten03a,luchinsky,peskin}). As such, they offer a complementary perspective to previously studied double vector or double pseudoscalar channels and serve as sensitive tests of relativistic bound-state effects built in into the Bethe–Salpeter amplitudes. 

The BSE–CIA framework is especially suitable for such processes because:
(i) it yields relativistic 4D hadron–quark vertices with full Dirac structure,
(ii) these hadron-quark vertices are directly connected to mass spectral equations (through the 3D Salpeter equations), thereby ensuring consistency between low-energy spectroscopy and high-energy production, and (iii) the 3D reduction of amplitudes still retains Lorentz covariance due to  their dependence on the internal momentum variable $\hat{q}^2$, which is a Lorentz-invariant variable,
allowing transition amplitudes to be calculated over the entire time-like domain. This approach has successfully described both low-energy observables— mass spectra, leptonic decay widths, radiative $M1/E1$ transitions \cite{lmenew18,eshete19,vaishali21a,bhatnagar20,vaishali21,vaishali25}—and high-energy exclusive cross sections \cite{monika23,bhatnagar24,bhatnagar25,bhatnagar25a}, without changing the common set of input parameters, that were fixed through mass spectrum  of ground and excited states of charmonia of various $J^{PC}$. 

In this work, we calculate the leading-order amplitudes and total cross sections for these processes $(O(\alpha^4))$ within the BSE framework, employing the interaction kernel, model parameters, and relativistic wave functions that were fixed from spectroscopy and successfully applied in our earlier studies \cite{bhatnagar24,bhatnagar25}. Since the relevant energy scale is large, contributions from higher-order QCD and pure electromagnetic corrections are expected to be subleading and are beyond the scope of the present leading order study in this paper. Our results thus provide predictions for these C-conserving double charmonium channels without introducing additional parameters beyond those fixed from spectroscopy, which may be relevant for future experimental investigations at B-factories and next-generation $e^- e^+$ colliders. And since no experimental data are currently available, our results constitute baseline theoretical predictions for these channels. 

Unlike previously studied double-vector or double-pseudoscalar channels, the present processes are free from photon-fragmentation enhancements and therefore expose the underlying quark-rearrangement dynamics more directly. In particular, the identical double P-wave channel $h_c+h_c$ provides a rare opportunity to study symmetry-driven interference effects and the sensitivity of production amplitudes to the momentum-space structure of P-wave Bethe–Salpeter wavefunctions. These features make the present channels qualitatively different from commonly studied double-charmonium processes and offer new insight into relativistic quarkonium production mechanisms.

The paper is organized as follows. In Section~2, we summarize the BSE framework for $Q\bar{Q}$ system. Section~3 provides the details
of the amplitude and cross section calculation for $e^- e^+\rightarrow\gamma^*\gamma^*\rightarrow \chi_{c1}+\eta_c$. Section~4 presents calculation of the process, $e^- e^+\rightarrow\gamma^*\gamma^*\rightarrow h_c+h_c$. In Section 4.1, we study the sensitivity of our predicted cross sections of both $\chi_{c1}+\eta_c$ and $h_c+h_c$ channels to input parameters of the framework. Finally, Section~5 is relegated to Discussions.

\section{Bethe--Salpeter equation framework}

We employ a relativistic Bethe--Salpeter equation (BSE) framework to describe
$Q\bar Q$ quarkonium bound states. The 4D BSE for a meson with total momentum $P$
and internal relative momentum $q$ is given by
\begin{equation}
\Psi(P,q)=S_F(p_1)\, i\!\int\!\frac{d^4q'}{(2\pi)^4}\,
K(q,q')\,\Psi(P,q')\,S_F(p_2),
\end{equation}
where $S_F(\pm p_{1,2})$ are the quark and antiquark propagators and $K(q,q')$
is the interaction kernel.

The equation is reduced to a covariant 3D form using the
Covariant Instantaneous Ansatz (CIA), wherein the kernel depends only on the
transverse component
\(\hat q_\mu = q_\mu - (q\!\cdot\!P/P^2) P_\mu\)\cite{mitra92,wang16,he19,he21,wang22},
satisfying $\hat q\!\cdot\!P=0$.
The 4D BS wave function can then be written as
\begin{equation}
\Psi(P,q)=S_1(p_1)\,\Gamma(\hat q)\,S_2(-p_2),
\end{equation}
where $\Gamma(\hat q)$ is the hadron--quark vertex function, obtained from the
3D BS wave function $\psi(\hat q)$,
\begin{equation}
\Gamma(\hat q)=\int\!\frac{d^3\hat q'}{(2\pi)^3}\,
K(\hat q,\hat q')\,\psi(\hat q').
\end{equation}

Following a sequence of steps we arrive at the 3D Salpeter equations that lead to the mass spectral equations, whose solutions not only lead to the N-dependent mass spectum, but also the 3D radial wave functions for ground and excited states of various $J^{PC}$(for details, see \cite{lmenew18,vaishali21a,eshete19,bhatnagar24,bhatnagar25}). Both the vertex function and 3D radial Salpeter wave functions depend on the Lorentz-invariant variable \(\hat q^2=q^2-(q\!\cdot\!P)^2/P^2\)\cite{mitra92,eshete19,bhatnagar20,vaishali21,bhatnagar24},
ensuring covariance beyond the hadron rest frame. These 3D radial wave functions of quarkonia are subsequently used in various transition amplitude calculations. Thus, this framework has been successfully applied to spectroscopy and decay
processes (leptonic\cite{eshete19,hluf16,vaishali21a}, radiative{\cite{bhatnagar20,vaishali21,vaishali25}, and two-photon decays\cite{lmenew18}), as well as to exclusive
quarkonium production\cite{bhatnagar24,monika23,bhatnagar25,bhatnagar25a} in $e^-e^+$ annihilation at
$\sqrt{s}=10.6$~GeV, with results in good agreement with experiment and other
models.
The present work extends these applications to
$\chi_{c1}+\eta_c$ and $h_c+h_c$ production within the same unified framework,
without introducing new parameters.

However in our quark propagators in BSE, we use constituent quark mass, though in principle, the dressed quark propagators
arise as the solution of the gap equation \cite{bhagwat06,roberts96,roberts03, roberts03a,roberts11}, which is characterized by a momentum-dependent mass function, $m(p)$. It has
been shown that constituent quark masses in the propagators of heavy (c,b) quarks is a good approximation \cite{chang10,wang05}, and provides
a rationale for constituent quark masses employed for heavy quarks in potential models, and the constituent quarks masses in the
propagators for heavy (c,b) quarks have also been employed in recent BSE calculations in \cite{chang10,shi20,wang05,wang22,vaishali21,wang16,he21,bhatnagar24,bhatnagar25}. Thus in the
present work, we will use constituent quark masses for c,b quarks in propagators, though for future work, we do intend to use
dressed quark propagators, for which the model will have to be modified.

As regards the interaction kernel $K(\hat{q}, \hat{q}')$, it is one-gluon-exchange like as regards the colour and spin dependence, and thus has a general structure \cite{eshete19,bhatnagar24},

\begin{eqnarray}
&&\nonumber K(\hat{q},\hat{q}')=(\frac{1}{2}\lambda_1.\frac{1}{2}\lambda_2)\gamma_{\mu}\times \gamma_{\mu} V(\hat{q}, \hat{q}')\\&&
\nonumber V(\hat{q},\hat{q}')= \frac{4\pi\alpha_s(M^2)}{(\hat{q}-\hat{q}')^2}
 +\frac{3}{4}\omega^2_{q\bar q}\int d^3r\bigg(\kappa r^2-\frac{C_0}{\omega_0^2}\bigg)e^{i(\hat q-\hat q').\vec r}\equiv V_{OGE}(\hat{q},\hat{q}')+V_{conf.}(\hat{q},\hat{q}'),\\&&
\nonumber \omega_{q\bar{q}}^2=2m\omega_0^2\alpha_s(M^2),\\&&
\alpha_s(M^2)=\frac{12\pi}{(33-2f)}[Log(M^2/\Lambda^2)]^{-1},
\end{eqnarray}

where the input parameters are: the flavour independent spring constant, $\omega_0=0.22$ GeV, charmed quark mass, $m=1.490 $GeV, QCD length scale, $\Lambda=0.250$ GeV, and dimensionless constants, $C_0=0.69$, $A_0=0.01$\cite{eshete19,bhatnagar24}. The input values used here are fixed earlier by fitting the mass spectra of
$0^{++},\,1^{--},\,0^{-+},\,1^{+-}$, and $1^{++}$ quarkonium states\cite{eshete19,vaishali21a}. Here, $\omega_{q\bar{q}}^2$ represents the flavor-dependent spring constant of the confining part of the BSE kernel. While not identical to the QCD confinement scale $\Lambda$, it effectively controls the strength of confinement in the model and determines the spatial scale of the bound-state wave functions.

\section{Cross section for $\chi_{c1} + \eta_c$ production in $e^- e^+$ annihilation}
The process $e^{-}e^{+} \rightarrow \gamma^{\ast}\gamma^{\ast} \rightarrow \chi_{c1} + \eta_{c}$
proceeds exclusively through quark–rearrangement (non–fragmentation) diagrams shown in Fig.1, where $H, H'=\chi_{c1}, \eta_c$ respectively. This restriction arises from 
the \(C\)-parity properties of the final states: while $\eta_{c}$ has  C = +1, 
the axial–vector meson $\chi_{c1}(1^{++})$ also carries $C = +1$. Since a single 
virtual photon has $C = -1$, a photon–fragmentation topology, in which each meson 
is produced independently from a single photon, is forbidden by C-parity conservation. 
Consequently, only the double–virtual–photon rearrangement diagrams contribute to this 
process.

\begin{figure}[ht!]
 \centering
 \includegraphics[width=10cm,height=5cm]{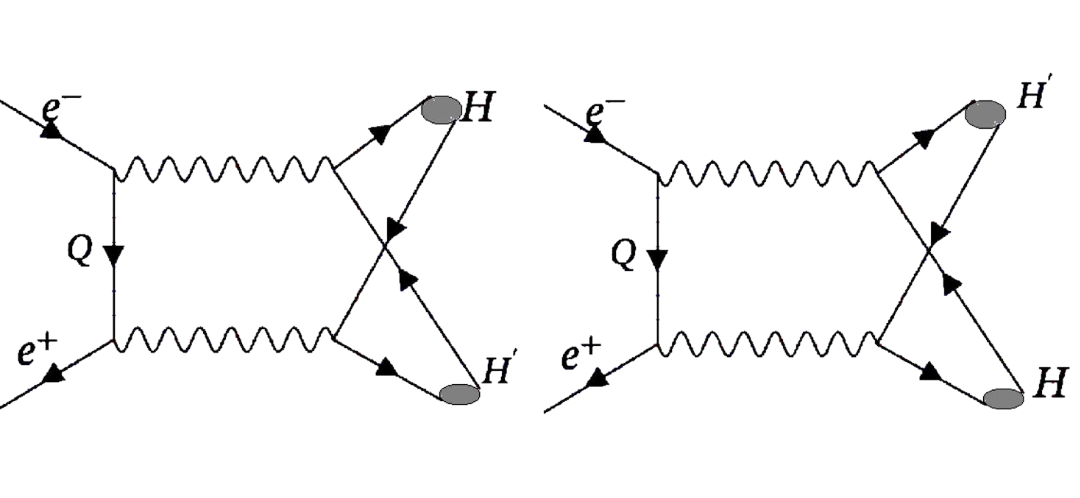}
 \caption{Feynman diagrams involved in $e^- e^+\rightarrow \gamma^*\gamma^*\rightarrow H+H'$ at leading order.}
 \end{figure}

We take $P,q, \epsilon$ as the external momentum, internal momentum and the polarization vector of the outgoing axial meson $\chi_{c1}$, and  $P',q'$ as the external and internal momenta of pseudoscalar meson, $\eta_c$, while $k$ and $k'$ as the momenta of the two virtual photons involved in the process. For the two quark–rearrangement diagrams, we adopt the heavy–quark approximation used in previous charmonium studies\cite{bhatnagar25}, where internal relative-momentum effects are suppressed by $\sim O(v^2)$, implying $\hat{q}^2/m^2 << 1$. In this limit, the photon momenta are effectively insensitive to the internal quark motion, and the photon virtualities satisfy $k^2\approx k'^2\approx s/4$. The product of the photon propagators then satisfies, $1/(k^2k'^2)\approx 16/s^2$. This is further justified since the photon virtualities are dominated by the external kinematics at $\sqrt{s}>> mv$. With this, the amplitudes for the two diagrams (Direct and Exchange) of the process $ e^{-}e^{+} \rightarrow \chi_{c1} + \eta_{c}$ can be written as:

\begin{equation}
M_{fi}|_D = \frac{16 e^2 e_Q^2}{s^2}
\left[ \bar{v}_2 \gamma_\nu 
\left( \frac{-i\not{Q} + m_e}{Q^2 + m_e^2} \right) 
\gamma_\mu u_1 \right] \int \frac{d^4 q}{(2\pi)^4} 
\int \frac{d^4 q'}{(2\pi)^4} 
\text{Tr}[ \gamma_\mu \bar{\psi}_A(P,q) \gamma_\nu \bar{\psi}_P(P',q')]
\end{equation}

and 
\begin{equation}
M_{fi}|_X= \frac{16 e^2 e_Q^2}{s^2}
\left[ \bar{v}_2 \gamma_\nu 
\left( \frac{-i\not{Q} + m_e}{Q^2 + m_e^2} \right) 
\gamma_\mu u_1 \right] \int \frac{d^4 q}{(2\pi)^3} 
\int \frac{d^4 q'}{(2\pi)^4} 
\text{Tr}\left[ \gamma_\mu \bar{\psi}_P(P',q') \gamma_\nu \bar{\psi}_A(P,q)\right],\\
\end{equation}

where the amplitude for the rearrangement diagram that involves the exchange of the final state mesons in the process $e^{-}e^{+} \rightarrow \chi_{c1} + \eta_{c}$ can again be obtained from the above amplitude through the exchange of Lorentz indices as ${M}^{(X)}_{fi} = {M}^{(D)}_{fi}|_{\mu \leftrightarrow \nu}$. To reduce the above amplitudes to their 3D forms, we make use of the Covariant Instantaneous Ansatz (CIA), under which the 4D volume elements of the internal momenta of the two produced hadrons can be written as $d^{4}q = d^{3}\hat{q} M d\sigma$ and $d^{4}q' = d^{3}\hat{q}' M'd\sigma'$. We then integrate over the longitudinal components $M d\sigma$ and $M'd\sigma'$ of the internal momenta $q$ and $q'$ of the two hadrons to obtain $\frac{1}{2\pi i}\int M d\sigma \Psi(P,q) = \bar{\Psi}(\hat{q})$ and $\frac{1}{2\pi i}\int M' d\sigma' \Psi(P',q') = \bar{\Psi}(\hat{q}')$, which are the corresponding 3D Bethe–Salpeter wave functions used to express the full amplitudes for the process $e^{-}e^{+} \rightarrow \chi_{c1} + \eta_{c}$ in a manifestly 3D form.

Thus, for the direct amplitude of photon fragmentation diagrams, we reduce the amplitude to its 3D form as
\begin{equation}
M_{fi}|_D = \frac{16 e^2 e_Q^2}{s^2}
\left[ \bar{v}_2 \gamma_\nu 
\left( \frac{-i\not{Q} + m_e}{Q^2 + m_e^2} \right) 
\gamma_\mu u_1 \right] \int \frac{d^3\hat{q}}{(2\pi)^3} 
\int \frac{d^3 \hat{q'}}{(2\pi)^3} 
\text{Tr}[ \gamma_\mu \bar{\psi}_A(\hat{q}) \gamma_\nu \bar{\psi}_P(\hat{q'})]
\end{equation}

The amplitude for the exchange process can be similarly written as:
\begin{equation}
M_{fi}|_X = \frac{16 e^2 e_Q^2}{s^2}
\left[ \bar{v}_2 \gamma_\nu 
\left( \frac{-i\not{Q} + m_e}{Q^2 + m_e^2} \right) 
\gamma_\mu u_1 \right] \int \frac{d^3\hat{q}}{(2\pi)^3} 
\int \frac{d^3 \hat{q'}}{(2\pi)^3} 
\text{Tr}[ \gamma_\mu \bar{\psi}_P(\hat{q}') \gamma_\nu \bar{\psi}_A(\hat{q})],
\end{equation}

and the 3D forms of adjoint Bethe-Salpeter wave functions of Dimensions $M$ for $\chi_{c1}$ and $\eta_c$ obtained by 3D reduction of the full 4D BS wave functions\cite{smith69,alkofer02} with all the Dirac structures are expressed as \cite{eshete19,monika23}:

\begin{eqnarray}
&&\nonumber \bar{\psi}_A(\hat{q})= N_A \left[ iM\not{\epsilon} + \not{P}\not{\epsilon} + \frac{2i\not{\hat{q}}\not{P}\not{\epsilon}}{M} \right] \gamma_5 \phi_A(\hat{q}),\\&&
\bar{\psi}_P(\hat{q}') = N_P \gamma_5 \left[ M' - i\not{P}' + \frac{2\not{\hat{q}'}\not{P}'}{M'} \right] \phi_P(\hat{q}').
\end{eqnarray}

where the radial wave functions for $\chi_{c1}(nP)$\cite{vaishali21a} obtained through solutions of its mass spectral equation are:
\begin{eqnarray}
&&\nonumber \phi_A(\hat{q},1P)=\sqrt{\frac{2}{3}} \frac{1}{\pi^{\frac{3}{4}}\beta_{A}^{\frac{5}{2}}}\hat{q}e^{-\frac{\hat{q}^2}{2\beta_{A}^2}}\\&&
\phi_A(2P,\hat{q})=\sqrt{\frac{5}{3}} \frac{1}{\pi^{\frac{3}{4}}\beta_{A}^{\frac{5}{2}}}\hat{q}(1-\frac{2}{5\beta_{A}^2}\hat{q}^2)e^{-\frac{\hat{q}^2}{2\beta_{A}^2}},
\end{eqnarray}

where its inverse range parameter $\beta_A=(\frac{2}{3}M\omega_{q\bar{q}}^2)^{1/4}$ \cite{monika23} is derived from the mass spectral equation for $\chi_{c1}(nP)$, with $\omega_{q\bar{q}}^2=M\omega_{0}^2\alpha_s(M^2)$\cite{eshete19,bhatnagar25} being the flavour dependent spring constant. The corresponding radial wave functions for $\eta_c(nS)$\cite{eshete19} are:
\begin{eqnarray}
&&\nonumber \phi_P(\hat{q},1S)= \frac{1}{\pi^{\frac{3}{4}}\beta_{P}^{\frac{3}{2}}}e^{-\frac{\hat{q}^2}{2\beta_{P}^2}}\\&&
\phi_P(\hat{q},2S)=\sqrt{\frac{3}{2}} \frac{1}{\pi^{\frac{3}{4}}\beta_{P}^{\frac{3}{2}}}(1-\frac{2}{3\beta_{P}^2}\hat{q}^2)e^{-\frac{\hat{q}^2}{2\beta_{P}^2}}.
\end{eqnarray}

where its inverse range parameter is: $\beta_P=(M\omega_{q\bar{q}}^2)^{1/4}$\cite{monika23,eshete19}. And $N_A$, and $N_P$ are the 4D BS normalizers that are obtained through the current conservation condition,

\begin{equation}\label{46}
2iP_\mu=\int \frac{d^{4}q}{(2\pi)^{4}}
\mbox{Tr}\left\{\overline{\Psi}(P,q)\left[\frac{\partial}{\partial
P_\mu}S_{F}^{-1}(p_1)\right]\Psi(P,q)S_{F}^{-1}(-p_2)\right\} +
(1\rightleftharpoons2),
\end{equation}

where, $\Psi(P,q)$ is the 4D BS wave function, while $\overline{\Psi}(P,q)=\gamma_4 \Psi(P,q)^{\dag}\gamma_4$ is the adjoint BS wave function. The above equation is reduced to 3D form by integrating the RHS over the longitudinal component $Md\sigma$, which reduces $\Psi_A(P,q)$ (and $\bar{\Psi}_P(P,q)$) to their 3D forms $\psi_A(\hat{q})$ (and $\bar{\psi}_P(P,q)$), whose algebraic forms are in Eqs.(9). Evaluating the trace over the gamma matrices, and multiplying both sides by $P_{\mu}/(-M^2)$, we then extract the BS normalizers:

\begin{eqnarray}
&&\nonumber N_{A}^{-2} = \frac{4M^2}{m} \int \frac{d^3\hat{q}}{(2\pi)^3} \phi_{A}^{2}(\hat{q}),\\&&
 N_{P}^{-2}=\frac{16M}{m}\int \frac{d^3\hat{q}}{(2\pi)^3}\frac{\hat{q}^2}{\omega} \phi_{P}^{2}(\hat{q}).
\end{eqnarray}

whose numerical values for $N_A$ and $N_P$ for the process studied are listed in table 2. The total amplitude $M_{fi}$ after trace evaluation over the gamma matrices, and making use of the fact that $P'.\epsilon=0$ in the center of mass frame, can be written as:

\begin{eqnarray}
&&\nonumber M_{fi}|_D = N_P N_A \frac{16 e^2 e_Q^2}{s^2}
\int \frac{d^3\hat{q}}{(2\pi)^3} \phi_A(\hat{q})
\int \frac{d^3\hat{q}'}{(2\pi)^3} \phi_P(\hat{q'})
\Big\{4M' \Big[ \bar{v}_2 \not{\epsilon} 
\Big( \frac{-i\not{Q} + m_e}{Q^2 + m_e^2} \Big)
\not{P} u_1 \Big]\\&&
\nonumber \quad -4M' \Big[ \bar{v}_2 \not{P} 
\Big( \frac{-i\not{Q} + m_e}{Q^2 + m_e^2} \Big)
\not{\epsilon} u_1 \Big]+4M \Big[ \bar{v}_2 \not{P'} 
\Big( \frac{-i\not{Q} + m_e}{Q^2 + m_e^2} \Big)
\not{\epsilon} u_1 \Big]+4M \Big[ \bar{v}_2 \not{\epsilon} 
\Big( \frac{-i\not{Q} + m_e}{Q^2 + m_e^2} \Big)
\not{P'} u_1 \Big]\Big\}\\
\end{eqnarray}

The above expression can be further compactified by introducing $G_1$ (for axial vector meson $\chi_{c1}$), and $G_2$ (for pseudoscalar meson $\eta_c$) as 3D integrals involving the two outgoing  mesons, defined as:
\begin{eqnarray}
&&\nonumber G_1=\int \frac{d^3\hat{q}}{(2\pi)^3}\phi_A(\hat{q})\\&&
G_2=\int \frac{d^3\hat{q}'}{(2\pi)^3}\phi_P(\hat{q}'),
\end{eqnarray}

where in the above integrals, the unnormalized forms of the 3D radial wave functions $\phi_A(\hat{q})$ and $\phi_P(\hat{q})$  are taken from Eqs. (10-11), since to calculate the transition amplitude, $M_{fi}$, we use the 4D BS normalizers $N_A$ and $N_P$ that are obtained through the covariant current condition Eq.(12). The 4D BS normalizer already encodes the full normalization of the bound-state wave function. Imposing an additional 3D normalization on $\phi_A$ and $\phi_P$  would therefore constitute a double-counting of the wave function normalization. This is consistent with the normalization convention adopted throughout the present work and in our earlier studies \cite{bhatnagar20, vaishali21, bhatnagar24,monika23, bhatnagar25, bhatnagar25a}. 

Further here, we have made use of the fact that $\hat{q}$ is an effective 3D variable and is a directional vector, due to which all terms in Trace with $\hat{q}$ become zero after the 3D integration over $d^3\hat{q}$. Further the terms involving $P'.\epsilon$ vanish. This is due to the fact that $P.\epsilon=\overrightarrow{P}.\overrightarrow{\epsilon}=0$, since momentum, $P=(\overrightarrow{P}, iE)$ of axial meson is always orthogonal to its polarization vector, $\epsilon=(\overrightarrow{\epsilon},i0)$. And in center of mass frame, $\overrightarrow{P}'=-\overrightarrow{P}$, which implies that $\overrightarrow{P}'.\overrightarrow{\epsilon}=P'.\epsilon=0$.

The first and second term contributes equally in the $|M_{fi}|^2$ and the same is true as regards the third term and fourth term in Eq.(14). The Direct amplitude can then be written as:

\begin{eqnarray}
&&\nonumber M_{fi}|_D = N_P N_A \frac{16 e^2 e_Q^2}{s^2}G_1G_2
\Big\{4M' \Big[ \bar{v}_2 \not{\epsilon} 
\Big( \frac{-i\not{Q} + m_e}{Q^2 + m_e^2} \Big)
\not{P} u_1 \Big] -4M' \Big[ \bar{v}_2 \not{P} 
\Big( \frac{-i\not{Q} + m_e}{Q^2 + m_e^2} \Big)
\not{\epsilon} u_1 \Big]\\&&
~~~~~~~~~~~~~~~~~~~~~~~~~~~~~~~~~~~~~~~~~~~~~+4M \Big[ \bar{v}_2 \not{P'} 
\Big( \frac{-i\not{Q} + m_e}{Q^2 + m_e^2} \Big)
\not{\epsilon} u_1 \Big]+4M \Big[ \bar{v}_2 \not{\epsilon} 
\Big( \frac{-i\not{Q} + m_e}{Q^2 + m_e^2} \Big)
\not{P'} u_1 \Big]\Big\},
\end{eqnarray}

while the exchange amplitude with the final states interchanged is:

\begin{eqnarray}
&&\nonumber M_{fi}|_X = N_P N_A \frac{16 e^2 e_Q^2}{s^2}G_1G_2
\Big\{-4M' \Big[ \bar{v}_2 \not{\epsilon} 
\Big( \frac{-i\not{Q} + m_e}{Q^2 + m_e^2} \Big)
\not{P} u_1 \Big] +4M' \Big[ \bar{v}_2 \not{P} 
\Big( \frac{-i\not{Q} + m_e}{Q^2 + m_e^2} \Big)
\not{\epsilon} u_1 \Big]\\&&
~~~~~~~~~~~~~~~~~~~~~~~~~~~~~~~~~~~~~~~~~~~~~+4M \Big[ \bar{v}_2 \not{P'} 
\Big( \frac{-i\not{Q} + m_e}{Q^2 + m_e^2} \Big)
\not{\epsilon} u_1 \Big]+4M \Big[ \bar{v}_2 \not{\epsilon} 
\Big( \frac{-i\not{Q} + m_e}{Q^2 + m_e^2} \Big)
\not{P'} u_1 \Big]\Big\}.
\end{eqnarray}

The total amplitude would then be: $M_{fi}= M_{fi}|_D + M_{fi}|_X$ is written as:

\begin{equation}
 M_{fi} = N_P N_A \frac{16 e^2 e_Q^2}{s^2}G_1G_2
\Bigg(8M \Big[ \bar{v}_2 \not{P'} 
\Big( \frac{-i\not{Q} + m_e}{Q^2 + m_e^2} \Big)
\not{\epsilon} u_1 \Big]+8M \Big[ \bar{v}_2 \not{\epsilon} 
\Big( \frac{-i\not{Q} + m_e}{Q^2 + m_e^2} \Big)
\not{P'} u_1 \Big]\Bigg).
\end{equation}

We now evaluate the spin averaged invariant amplitude modulus squared, $|\bar{M}_{fi}|^2=\frac{1}{4}\sum_{s1,s2,\lambda}{M^\dag}_{fi}M_{fi}$, for which we must now average over the initial spin states of $e^-$ and $e^+$ and sum over the polarisation states of the final axial vector meson $\chi_{c1}$. We make use of the normalization relation, $\Sigma_{\lambda} \epsilon_{\mu}^{\lambda} \epsilon_{\nu}^{\lambda} = \frac{1}{3}(\delta_{\mu\nu} + \frac{P_{\mu}P_{\nu}}{M^2})$ \footnote{We have made use of the "Euclidean" notation-treating time part as the fourth component of a four-vector, where our metric is $\delta_{\mu\nu}$} for the polarisation vector of the axial meson.

As regards the kinematics of the process is concerned, one can note that $\overline{p}_1+\overline{p}_2=k +k'=P+P'$. Also, since $k=\overline{p}_1-Q$, and $k'=Q+\overline{p}_2$, we can obtain the momentum relation, $\bar{p}_1-\bar{p}_2=2Q+k-k'\approx 2Q$. And in center of mass frame, where $\bar{p}_1=(\overrightarrow{p},iE)$, and $\bar{p}_2=(-\overrightarrow{p},iE)$, we can write, $Q=(\overrightarrow{p},i0)$. Thus we can obtain, 

\begin{equation}
Q^2=-{m_e}^2 +\frac{s}{4}. 
\end{equation}

Taking $\theta$ to be the angle between the incident beam direction and $\chi_{c1}$ (or $h_c$ in Section 4), the dot products of various momenta in the center of mass frame can be expressed as: ~$\bar{p_1}.\bar{p_2}=-\frac{s}{2}$,~ $\bar{p}_1.P=-\frac{s}{4}(1+Cos\theta)=\bar{p}_2.P'$, ~ $\bar{p}_1.P'=\frac{s}{4}(-1+Cos\theta)=\bar{p}_2.P$,~~$Q.P=\frac{s}{4}Cos\theta$, ~~ $Q.P'=-\frac{s}{4}Cos\theta$. Similarly,
$\bar{p_1}.Q=-\bar{p_2}.Q  \sim \frac{s}{4}$,~ ~$P.P'=-\frac{s}{2}+M^2$,~~ $P.I=0$,~~ $P'.I=0$. And $P.\epsilon=P'.\epsilon'=0$,~ ~ $P.\epsilon'=P'.\epsilon=0$. Similarly, the dot products: $p_1.I=p_2.I'=\frac{\sqrt{s}}{2}Sin\theta$,~~ $p_1.I'=p_2.I=-\frac{\sqrt{s}}{2}Sin\theta$, ~~ $I.Q=\frac{\sqrt{s}}{2}Sin\theta$, and $I'.Q=-\frac{\sqrt{s}}{2}Sin\theta$. Here $I=\frac{\hat{q}}{|\hat{q}|}$, and $I'=\frac{\hat{q}'}{|\hat{q}'|}$ are unit vectors along the direction of $\hat{q}$ and $\hat{q}'$ respectively, while $|\hat{q}|=\sqrt{q^2-\frac{(q.P)^2}{P^2}}$ and $|\hat{q}'|=\sqrt{{q'}^2-\frac{(q'.P')^2}{P'^2}}$ are Lorentz-invariant variables. Also, $(\bar{p_1}+\bar{p_2})^2=-s$. These kinematical relations are valid for the quark rearrangement diagrams applicable for these transitions. We obtain

\begin{eqnarray}
\left|\mathcal{M}_{fi}\right|^2 &=&
\frac{e^4\, e_Q^4 \cdot 16^2\, G_1^2\, G_1^{\prime\,2}}
     {24\, s^4 \left(Q^2 + m_e^2\right)^2}
\Bigg[
- \cos^4\theta \left(s_1^2 - 4M^2\right)^2 \left(s_1^2 - 4m_e^2\right)^2
\nonumber \\
&& +\; \cos^2\theta \; s_1^2 \left(s_1^2 - 2M_p^2\right)
\left(s_1^2 - 4M^2\right)\left(s_1^2 - 4m_e^2\right)
\nonumber \\
&& +\; M^2 \Bigg(
- \cos^3\theta \left(s_1^2 - 4M^2\right)^{3/2}\left(s_1^2 - 4m_e^2\right)^{3/2}
\nonumber \\
&& \qquad +\; 2\cos^2\theta \; s_1^2
\sqrt{s_1^2 - 4M^2}\left(s_1^2 - 4m_e^2\right)\sqrt{s_1^2 - 4M_p^2}
\nonumber \\
&& \qquad +\; 12\,m_e^2\cos^2\theta
\left(s_1^2 - 4M^2\right)\left(s_1^2 - 4m_e^2\right)
\nonumber \\
&& \qquad -\; 2\cos\theta \; s_1^2\left(s_1^2 - 2m_e^2\right)
\sqrt{\left(s_1^2 - 4M^2\right)\left(s_1^2 - 4m_e^2\right)}
\nonumber \\
&& \qquad +\; s_1^2\left(s_1^2 - 4m_e^2\right)\left(s_1^2 - 4M_p^2\right)
\Bigg) \Bigg]
\end{eqnarray}

Here $\alpha=\frac{e^2}{4\pi}$ is the QED coupling constant, while, $N_A$ and $N_P$ are the BS normalizers of axial and pseudoscalar mesons and are evaluated through the current conservation condition, Eq.(12). The total cross section for the process is given as,

\begin{equation}
\sigma=\frac{1}{32\pi^2 s^{3/2}}|\vec{P'}|\int d\Omega' |\bar{M}_{fi}|^2,
\end{equation}

with $|\vec{P'}|=\sqrt{\frac{1}{s}[s-(M+M')^2][s-(M-M')^2]}$ being the momenta of either of the final particles. It is observed that the dimensions of the following quantities are as follows: $G_1\sim M^4$, $G_2 \sim M^3$, $N_A\sim M^{-3}$, $N_P\sim M^{-2}$, $|\bar{M}_{fi}|^2\sim M^0$, due to which the cross section, $\sigma\sim M^{-2}$. The numerical values of cross sections for $e^- e^+\rightarrow \chi_{c1}(nP) + \eta_c (nS)$ (with n=1,2) in our work at $\sqrt{s}$=10.6 GeV along with the results of NRQCD\cite{braaten03} are given in Table 1.

\begin{table}[hhhhh]
\centering
\caption{Sensitivity of the predicted cross sections (in fb) for the process, $e^- e^+\rightarrow \gamma^*\gamma^*\rightarrow \chi_{c1}(nP)+\eta_c(nS)$  to independent variations of the confinement parameter $\omega_0$, QCD scale $\Lambda$, and the charm-quark mass $m$.}
\vspace{0.2cm}

\begin{tabular}{|c|c|c|c|c|c|c|}
\hline

$m_c$ (GeV) &$\omega_0$ (GeV) &$\Lambda$ (GeV) &$\sigma(1P+1S)$ &$\sigma(1P+2S)$ &$\sigma(2P+1S)$ &$\sigma(2P+2S)$\\

\hline

1.490(base) & 0.220  & 0.250 &$1.315\times 10^{-3}$&$0.799\times 10^{-3}$&$0.992\times 10^{-3}$& $0.606\times 10^{-3}$\\

\hline

1.490 & 0.210 & 0.250 &$1.193\times 10^{-3}$ & $0.722\times 10^{-3}$&$0.899\times 10^{-3}$& $0.547\times 10^{-3}$\\

\hline

1.490 & 0.230 & 0.250&$1.440\times 10^{-3}$ &$0.881\times 10^{-3}$ &$1.089\times 10^{-3}$&$0.668\times 10^{-3}$\\

\hline

1.490 & 0.220 & 0.238 &$1.288\times 10^{-3}$ &$0.783\times 10^{-3}$&$0.972\times 10^{-3}$&$0.594\times 10^{-3}$\\

\hline

1.490 & 0.220 & 0.261 &$1.340\times 10^{-3}$& $0.813\times 10^{-3}$&$1.010\times 10^{-3}$&$0.616\times 10^{-3}$
\\

\hline

1.470 & 0.220 & 0.250 &$1.261\times 10^{-3}$&$0.767\times 10^{-3}$&$0.951\times 10^{-3}$ &$0.582\times 10^{-3}$\\

\hline

1.510 & 0.220 & 0.250 &$1.371\times 10^{-3}$&$0.832\times 10^{-3}$&$1.033\times 10^{-3}$ &$0.630\times 10^{-3}$\\

\hline
\end{tabular}

\label{tab:decay-widths_sensitivity_total}
\end{table}

\begin{table}[hhhhh]
  \begin{center}
  \begin{tabular}{p{4cm} p{3cm} p{3.5cm}p{3.5cm}p{3.5cm} }
  \hline
 Process                                         &$N_A$ &$ G_1$   & $N_P$& $G_2$\\
  \hline
  $e^- e^+\rightarrow \chi_{c1}(1P) \eta_c(1S)$  & 3.930 &  0.0104&5.9596 &0.00885\\
  $e^- e^+\rightarrow \chi_{c1}(2P) \eta_c(2S)$  & 4.734 & -0.0081&4.6308&-0.00929 \\                      \\
     \hline
  \end{tabular}
\caption{Numerical values of BS normalizer $N_A$ (in $GeV^{-3}$)  and the 3D integral $G_1$ (in $GeV^4$ ) for $\chi_{c1}(nP)$, and the corresponding values of $N_P$ (in $GeV^{-2}$) and $G_2$ (in $GeV^3$) for $\eta_c(nS)$ employed in the process, $e^- e^+\rightarrow \chi_{c1}(nP)+ \eta_c(nS)$ at $\sqrt{s}=10.6 GeV$.}
\end{center}
\end{table}

Table 2 gives the values of the BS normalizer $N_A$/ $N_P$ and the 3D integrals, $G_1$/ $G_2$ for $\chi_{c1}$/ $\eta_c$ respectively. Here, the integrals, $G_1$ are in turn related to the leptonic decay constants, $f_A$ of the produced $\chi_{c1}$ mesons, due to the fact that the leptonic decay constant expression, $f_A=4\sqrt{3} N_A \int \frac{d^3\hat{q}}{(2\pi)^3}\phi_A(\hat{q})$\cite{vaishali21a}, and thus involves the product $N_{A} G_1$. And from Table 2, it can be checked that $N_{A} G_1$ for $\chi_{c1}(1P)$ is greater than that for $\chi_{c1}(2P)$- which is in conformity with the fact that the leptonic decay constant value, $f_A$ is greater for $\chi_{c1}(1P)$ than for $\chi_{c1}(2P)$, mainly due to the presence of a node in the wave function for 2P state (see \cite{vaishali21a}). The same holds for $\eta_c$ meson as well, where the product $N_PG_2$ (present in the expression for decay constant $f_P$\cite{eshete19}) for 1S state is greater than that for 2S state.

Now to assess the parameter-induced uncertainties in the predicted cross sections for the process $e^-e^+ \rightarrow \gamma^*\gamma^* \rightarrow \chi_{c1}(nP) + \eta_c(nS)$,
we independently varied the three model parameters: the confinement scale $\omega_0$, the QCD scale $\Lambda$, and the charm-quark mass m about their base values of $\omega_0 = 0.220$ GeV,
$\Lambda = 0.250$ GeV, $m=1.490$ GeV. The confinement parameter was varied over the range $\omega_0 \in [0.210, 0.230]$ GeV, the QCD scale over $\Lambda \in [0.238, 0.261]$ GeV, and the charm-quark mass over $m \in [1.470, 1.510]$ GeV, with the remaining two parameters held fixed in each case. The individual contributions to the uncertainty were combined in quadrature, yielding the results shown in Table 3, corresponding to a total relative parameter uncertainty of approximately $11\%$ in each case. A breakdown of the variance reveals a consistent hierarchy across all four processes: the confinement parameter $\omega_0$ is overwhelmingly dominant, accounting for approximately $82–85\% $ of the total variance with a relative sensitivity of $\sim10\%$, while the charm-quark mass m
 contributes a secondary but non-negligible $\sim 13–15\%$ of the variance at a relative sensitivity of $\sim 4\%$, and the QCD scale $\Lambda$
 alone contributes only $\sim 2–3\%$ of the variance at $\sim 2\%$.

It is important to note that, m exerts a larger independent influence on the cross sections than $\Lambda$
alone, a result that contrasts with the behavior seen in the radiative decay widths $\Gamma(H \to \gamma\gamma$), where the combined confinement sector dominates far more strongly. This difference is physically understood as arising from the fact that production cross sections depend on global wavefunction overlap integrals and meson-mass-dependent kinematics, making them relatively more sensitive to m. The results of cross sections with parameter induced uncertainties are reported in Table 3.

\begin{table}[hhhhh]
  \begin{center}
  \begin{tabular}{p{6cm} p{3.5cm} p{4cm} }
  \hline
 Process                                      & Present work& NRQCD\cite{braaten03}\\
  \hline
  $e^- e^+\rightarrow \chi_{c1}(1P) \eta_c(1S)$  & $1.315^{+0.139}_{-0.136}\times 10^{-3}$ & $(1.31\pm 0.29)\times 10^{-3}$ \\
  $e^- e^+\rightarrow \chi_{c1}(1P) \eta_c(2S)$ & $0.799^{+0.089}_{-0.085}\times 10^{-3}$ &  $(0.540\pm0.120)\times 10^{-3}$  \\                              
  $e^- e^+\rightarrow \chi_{c1}(2P) \eta_c(1S)$  & $0.992^{+0.107}_{-0.104}\times 10^{-3}$ &                                 \\ 
  $e^- e^+\rightarrow \chi_{c1}(2P) \eta_c(2S)$ & $0.606^{+0.067}_{-0.065}\times 10^{-3}$ &                                  \\                                               
     \hline
  \end{tabular}
\caption{Cross sections (in fb) at leading order (LO) for process, $e^- e^+\rightarrow \chi_{c1}(nP) \eta_c(nS)$ calculated in present work at $\sqrt{s}=10.6 GeV$ along with results of other models.}
\end{center}
\end{table}

\begin{figure}[ht!]
 \centering
 \includegraphics[width=10cm,height=5cm]{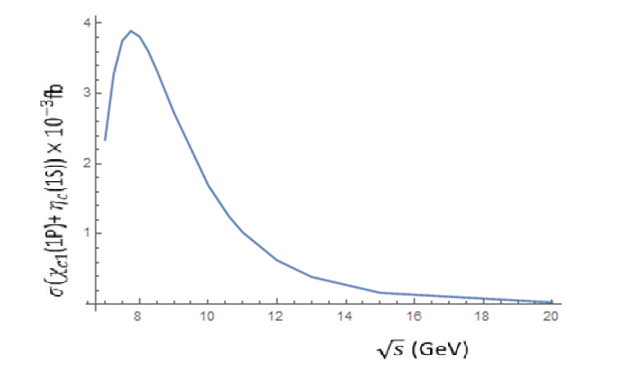}
 \caption{Plot of cross section for $e^- e^+\rightarrow \gamma^*\gamma^*\rightarrow \chi_{c1}(1P)+\eta_c(1S)$ (in fb) versus $\sqrt{s}$ (in GeV)}
 \end{figure}

Further from the cross section results it can be checked that the hierarchy is physically reasonable and consistent with expectations, and it can be justified quite cleanly within both BSE-type models and general QCD intuition. For the channels where NRQCD results are available (1P+1S and 1P+2S), both approaches agree that the ground-state channel dominates. Our BSE-CIA further predicts the ordering of excited channels.  We find that $\sigma[\chi_{c1}(1P)\eta_c(1S)]> \sigma[\chi_{c1}(2P)\eta_c(1S)]>\sigma[\chi_{c1}(1P)\eta_c(2S)]>\sigma[\chi_{c1}(2P)\eta_c(2S)]$. For the 1P+1S channel, our result agrees with NRQCD within combined uncertainties, while for the 1P+2S channel our BSE-CIA value exceeds the NRQCD prediction. Thus, although absolute magnitudes of our cross sections differ from NRQCD, the NRQCD pattern in \cite{braaten03} follows the same qualitative hierarchy: i.e. The channels involving ground states dominate, while the channels with one excited state are suppressed, and the channels with two excited states are most suppressed.

The plot of the cross section for $e^- e^+ \to \gamma^\ast \gamma^\ast \to \chi_{c1}(1P) + \eta_c(1S)$ versus $\sqrt{s}$ is given in Fig. 2. The cross section $\sigma$ scales with energy and shows a strong power law suppression $\sigma \sim 1/s^n$, which is consistent with the dimensional counting rules of exclusive processes. Although our calculation is performed in the Bethe–Salpeter framework, the observed energy dependence reflects the underlying short-distance structure of the hard process. At sufficiently large $s$, the scaling is governed by propagator behaviour and phase-space suppression, leading to a power-law falloff that is consistent with quark-counting expectations for exclusive processes involving a P-wave quarkonium.

\section{Cross section for $h_c + h_c$ production in $e^- e^+$ annihilation}
The process $e^{-}e^{+} \rightarrow \gamma^{\ast}\gamma^{\ast} \rightarrow h_c + h_c$ proceeds exclusively through quark–rearrangement (non–fragmentation) diagrams shown in Fig.1, where $H = H'=h_c$. The process $e^- e^+\rightarrow \gamma^*\gamma^*\rightarrow h_c+h_c$ proceeds dominantly through the production of two color-singlet $c\bar{c}$ pairs at leading order in $\alpha$. Since the virtual photons couple directly to color-singlet heavy-quark currents, color-octet configurations can only arise through additional hard-gluon exchanges and are therefore suppressed by higher powers of $\alpha_s$, and the heavy-quark velocity $v$. Moreover, unlike double vector production, the absence of photon fragmentation enhancement further suppresses any potential color-octet contributions. Consequently, the $h_c+h_c$ channel provides a particularly clean probe of color-singlet quarkonium dynamics in two-photon processes.

Although the $h_c$ state carries the same charge conjugation quantum number as the photon $(C=-1)$, a single photon fragmentation into $h_c$ is forbidden. This follows from the mismatch in parity and operator structure: while the electromagnetic current $\bar{c}\gamma_{\mu}c$ has $J^{PC}=1^{--}$, the $h_c$ meson is $1^{+-}$ state corresponding to a $^1P_1$ configuration, due to which the matrix element $<0|\bar{c}\gamma_{\mu}c|h_c>$ vanishes identically. Consequently, in the two-photon process $e^- e^+\rightarrow \gamma^*\gamma^*\rightarrow h_c + h_c$, neither photon can independently fragment into an $h_c$ meson. The production therefore proceeds only through non-fragmentation two-photon mechanisms, where both photons participate coherently in forming the final state and the required orbital angular momentum is generated dynamically. This leads to a strong suppression of the $h_c+h_c$  cross section compared to fragmentation-dominated channels such as $J/\Psi+J/\Psi$ consistent with the observed hierarchy among double-charmonium production processes.

For the first quark–rearrangement diagram, we follow the same heavy–quark approximation used in previous 
charmonium studies, treating the virtual–photon momenta as nearly independent of the internal relative 
momenta of the quark–antiquark pairs. Under this approximation, the photon virtualities satisfy 
\( k^{2} \simeq k'^{\,2} \simeq s/4 \), so that the product of the photon propagators simplifies as 
\( 1/(k^{2}k'^{\,2}) \approx 16/s^{2} \). With this, the amplitude for the two rearrangement diagrams of the process 
$ e^{-}e^{+} \rightarrow h_{c} + h_{c}$ can be written in a 4D form as in Eqs.(5-6), with both the adjoint wave functions being $\bar{\Psi}_A(P,q)$, where A stands for $h_c$ meson. Following the usual reduction to 3D form under Covariant Instantaneous Ansatz, we can express the 3D forms of the two amplitudes as: 

\begin{eqnarray}
&&\nonumber M_{fi}|_D = \frac{16 e^2 e_Q^2}{s^2}
\left[ \bar{v}_2 \gamma_\nu 
\left( \frac{-i\not{Q} + m_e}{Q^2 + m_e^2} \right) 
\gamma_\mu u_1 \right] \int \frac{d^3\hat{q}}{(2\pi)^3} 
\int \frac{d^3\hat{q}'}{(2\pi)^3} 
\text{Tr}\left[ \gamma_\mu \bar{\psi}_A(\hat{q}) \gamma_\nu \bar{\psi}_A(\hat{q}') \right],\\&&
 M_{fi}|_X = \frac{16 e^2 e_Q^2}{s^2}
\left[ \bar{v}_2 \gamma_\nu 
\left( \frac{-i\not{Q} + m_e}{Q^2 + m_e^2} \right) 
\gamma_\mu u_1 \right] \int \frac{d^3\hat{q}}{(2\pi)^3} 
\int \frac{d^3\hat{q}'}{(2\pi)^3} 
\text{Tr}\left[ \gamma_\nu \bar{\psi}_A(\hat{q}) \gamma_\mu \bar{\psi}_A(\hat{q}') \right]
\end{eqnarray}

with the  adjoint 3D BS wave functions of $h_c$ for the two final states with masses and momenta $M,P,q$ and $M',P',q'$ of dimension $M$ being given as\cite{monika23}:

\begin{eqnarray}
&&\nonumber \bar{\psi}_A(\hat{q})= -N_A(\hat{q}.\epsilon)\Big[1+i\frac{\not{P}}{M}+\frac{2\not{P}\not{\hat{q}}}{M^2}\Big]\gamma_5\phi_A(\hat{q})\\&&
\bar{\psi}_A(\hat{q'})= -N_A(\hat{q'}.\epsilon)\Big[1+i\frac{\not{P'}}{M'}+\frac{2\not{P'}\not{\hat{q}'}}{M'^2}\Big]\gamma_5\phi_A(\hat{q'}),
\end{eqnarray}
where $N_A$ is the 4D BS normalizer obtained through the current conservation condition as:

\begin{equation}
N_{A}^{-2}=\frac{2}{m^3}\int \frac{d^3\hat{q}}{(2\pi)^3}\frac{\hat{q}^2}{3}\phi_{A}^2(\hat{q})[M^2+2mM+2\omega^2].
\end{equation}

And the radial wave functions for $h_c(nP)$\cite{vaishali21a} obtained through solutions of its mass spectral equation are:
\begin{eqnarray}
&&\nonumber \phi_A(\hat{q},1P)=\sqrt{\frac{2}{3}} \frac{1}{\pi^{\frac{3}{4}}\beta_{A}^{\frac{5}{2}}}\hat{q}e^{-\frac{\hat{q}^2}{2\beta_{A}^2}}\\&&
\phi(2P,\hat{q})=\sqrt{\frac{5}{3}} \frac{1}{\pi^{\frac{3}{4}}\beta_{A}^{\frac{5}{2}}}\hat{q}(1-\frac{2}{5\beta_{A}^2}\hat{q}^2)e^{-\frac{\hat{q}^2}{2\beta_{A}^2}}.
\end{eqnarray}

The inverse range parameter $\beta_A=(\frac{4}{3}m\omega_{q\bar{q}}^2)^{1/4}$ \cite{monika23} is derived from the mass spectral equation for $h_c(nP)$, with $\omega_{q\bar{q}}^2=M\omega_{0}^2\alpha_s(M^2)$\cite{monika23,eshete19,bhatnagar25}.

We then obtain the 3D form of transition amplitudes for Direct and exchange processes that are identical, and written as:
\begin{eqnarray}
&&\nonumber M_{fi}|_D= M_{fi}|_X=\frac{e^2 e_{Q}^2N_A N_{A'}}{k^2k'^2}\frac{16}{MM'}F_1 F_2 (I.\epsilon)(I'.\epsilon')\times\\&&
\nonumber \bigg[-(I.I')P.P'[\bar{v}_2\gamma_{\mu}(\frac{-i{\not}Q+m_e}{Q^2+m_{e}^2})\gamma_{\mu}u_1]+(I.I')[\bar{v}_2{\not}P'(\frac{-i{\not}Q+m_e}{Q^2+m_{e}^2}){\not}P u_1]+\\&&
\nonumber (I.I')[\bar{v}_2{\not}P(\frac{-i{\not}Q+m_e}{Q^2+m_{e}^2}){\not}P' u_1]+(P.P')[\bar{v}_2{\not}I'(\frac{-i{\not}Q+m_e}{Q^2+m_{e}^2}){\not}I u_1]+\\&&
(P.P')[\bar{v}_2{\not}I(\frac{-i{\not}Q+m_e}{Q^2+m_{e}^2}){\not}I' u_1]\bigg],
\end{eqnarray}

where the 3D integrals, 

\begin{eqnarray}
&&\nonumber F_1=\int \frac{d^3\hat{q}}{(2\pi)^3}\hat{q}^2\phi_{A_1}(\hat{q})\\&&
F_2=\int \frac{d^3\hat{q}'}{(2\pi)^3}\hat{q}'^2\phi_{A_{2}}(\hat{q}'),
\end{eqnarray}

where in the above integrals, the unnormalized forms of the 3D radial wave functions $\phi_{A1}(\hat{q})$ and $\phi_{A2}(\hat{q})$  are taken from Eqs. (25), since to calculate the transition amplitude, $M_{fi}$, we use the 4D BS normalizer $N_A$,  that is obtained through the current condition Eq.(12). The 4D BS normalizer already encodes the full normalization of the bound-state wave function. Imposing an additional 3D normalization on $\phi_{A1}(\hat{q})$ and $\phi_{A2}(\hat{q})$ would therefore constitute a double-counting of the wave function normalization. This is consistent with the normalization convention adopted throughout the present work and in our earlier studies \cite{bhatnagar20, vaishali21, bhatnagar24,monika23, bhatnagar25, bhatnagar25a}.

and the BS normalizers $N_{A}$ and $N_{A'}$ are evaluated using the current conservation condition following a similar procedure listed after Eq.(12). Their numerical values are listed after Table 3.  As mentioned earlier,  $I=\frac{\hat{q}}{|\hat{q}|}$, and $I'=\frac{\hat{q}'}{|\hat{q}'|}$ are unit vectors along the direction of $\hat{q}$ and $\hat{q}'$ respectively, while $|\hat{q}|=\sqrt{q^2-\frac{(q.P)^2}{P^2}}$ and $|\hat{q}'|=\sqrt{{q'}^2-\frac{(q'.P')^2}{P'^2}}$ are Lorentz-invariant variables. Further,  $k^2 k'^2=\frac{s^2}{16}$. Here, we have made use of the fact that $\hat{q}$ is an effective 3D variable and is a directional vector, due to which all terms in trace with a single power of $\hat{q}$ become zero after the 3D integration over $d^3\hat{q}$. However the terms with $\hat{q}^2$ and $\hat{q}'^2$ survive after 3D integration, and are expressed as $F_1$ and $F_2$ given above. Further the terms involving $P'.\hat{q}$, and $P.\hat{q}'$ vanish in the center of mass frame. 

We now evaluate the spin averaged invariant amplitude modulus squared, $|\bar{M}_{fi}|^2=\frac{1}{4}\sum_{s1,s2,\lambda}{M^\dag}_{fi}M_{fi}$, for which we must now average over the initial spin states of $e^-$ and $e^+$ and sum over the polarisation states of the final axial vector meson $h_c$. We again make use of the normalization relation, $\Sigma_{\lambda} \epsilon_{\mu}^{\lambda} \epsilon_{\nu}^{\lambda} = \frac{1}{3}(\delta_{\mu\nu} + \frac{P_{\mu}P_{\nu}}{M^2})$ for the polarisation vector of the axial meson. $|\bar{M}_{fi}|^2$ is then written as:

\begin{eqnarray}
&&\nonumber |\bar{M}_{fi}|^{2}=(16)^4\frac{e^4 e_{Q}^4 F_{1}^2 F_{2}^2 N_{A}^2 N_{A'}^2}{M^2 M'^2 s^4}[TR];\\&&
\nonumber [TR]=\frac{1}{(m_{e}^2+Q^2)^2M^2M'^2} \Bigg[Cos^2\theta Q^2[M^4(8s-16m_{e}^2)+M^2s(16m_{e}^2-8s)+2s^2(s-2m_{e}^2)+\\&&
\nonumber Sin^2\theta \bigg[M^4\bigg(s^2(4-8Sin^2\theta)-8m_{e}^2s\bigg)+s^4(1-2Sin^2\theta)+M^2\bigg(8m_{e}^2s^2+s^3(8Sin^2\theta-4)\bigg)-2m_{e}^2s^3\bigg]-\\&&
\nonumber \frac{1}{2}Cos^6\theta (s-4M^2)^2(s-4m_{e}^2)^2+Cos^4\theta (s-4M^2)(s-4m_{e}^2)[M^2(7m_{e}^2-\frac{3}{2}s)+s(\frac{1}{2}s-3m_{e}^2)]+\\&&
\nonumber Cos^4\theta (s-4M^2)M^2(m_{e}^2-\frac{1}{2}s)(s-4m_{e}^2)+2sCos^3\theta Sin^2\theta(s-4M^2)(s-2M^2)(s-4m_{e}^2)+\\&&
\nonumber Cos^2\theta \bigg[M^4(-32m_{e}^4+16m_{e}^2s-2s^2)+M^2s(32m_{e}^4-16m_{e}^2s+2s^2)+s^2(-8m_{e}^2+4m_{e}^2s-\frac{1}{2}s^2)\bigg]+\\&&
Cos\theta Sin^2\theta \bigg[M^4(24m_{e}^2s-4s^2)+M^2(4s^3-24m_{e}^2s^2)+6m_{e}^2s^2-s^4\bigg]\Bigg]
\end{eqnarray}

where $[TR]$ is the trace over the gamma matrices. The dimensions of the following quantities are as follows: $F_1(=F_2)\sim M^6$, $N_A\sim M^{-3}$,  $|\bar{M}_{fi}|^2\sim M^0$, due to which the cross section, $\sigma\sim M^{-2}$.

\begin{table}[hhhhh]
\centering
\caption{Sensitivity of the predicted cross sections (in fb) for the process, $e^- e^+\rightarrow \gamma^*\gamma^*\rightarrow h_c(nP)+h_c(nP)$  to independent variations of the confinement parameter $\omega_0$, QCD scale $\Lambda$, and the charm-quark mass $m$.}
\vspace{0.2cm}

\begin{tabular}{|c|c|c|c|c|}
\hline

$m_c$ (GeV) &$\omega_0$ (GeV) &$\Lambda$ (GeV) &$\sigma(1P+1P)$ &$\sigma(2P+2P)$ \\

\hline

1.490(base) & 0.220  & 0.250 &$0.113\times 10^{-3}$&$0.298\times 10^{-3}$\\

\hline

1.490 & 0.210 & 0.250 &$0.092\times 10^{-3}$ & $0.239\times 10^{-3}$\\

\hline

1.490 & 0.230 & 0.250&$0.140\times 10^{-3}$ &$0.369\times 10^{-3}$ \\

\hline

1.490 & 0.220 & 0.238 &$0.108\times 10^{-3}$ &$0.286\times 10^{-3}$\\

\hline

1.490 & 0.220 & 0.261 &$0.117\times 10^{-3}$& $0.310\times 10^{-3}$\\

\hline

1.470 & 0.220 & 0.250 &$0.106\times 10^{-3}$&$0.279\times 10^{-3}$\\

\hline

1.510 & 0.220 & 0.250 &$0.120\times 10^{-3}$&$0.318\times 10^{-3}$\\

\hline
\end{tabular}

\label{tab:decay-widths_sensitivity_total}
\end{table}

The cross section is then evaluated from Eq.(21), and is given in Table below for $h_c(1P)+h_c(1P)$, and $h_c(2P)+h_c(2P)$ channels. The former is compared with the corresponding NRQCD prediction\cite{braaten03} for this channel.

\begin{table}[hhhhh]
  \begin{center}
  \begin{tabular}{p{6.3cm} p{3.1cm} p{3.5cm} }
  \hline
 Process                                      & BSE-CIA& NRQCD\cite{braaten03}\\
  \hline
  $e^- e^+\rightarrow h_c(1P) h_c(1P)$  & $0.113^{+0.028}_{-0.023}\times 10^{-3}$ & $(0.19\pm 0.02)\times 10^{-3}$ \\
  $e^- e^+\rightarrow h_c(2P) h_c(2P)$  & $0.298^{+0.075}_{-0.063}\times 10^{-3}$ &                                  \\
      \hline
  \end{tabular}
\caption{Cross sections (in fb) at leading order (LO) for process, $e^- e^+\rightarrow h_c h_c$ calculated in present work at $\sqrt{s}=10.6 GeV$ along with results of NRQCD.}
\end{center}
\end{table}

For $h_c(1P)+h_c(1P)$, numerical values of BS normalizer, $N_A=N_{A'}=13.664GeV^{-3}$, and the 3D integrals, $F_1=F_2=0.0149 GeV^6$, while for $h_c(2P)+h_c(2P)$, $N_A=12.775GeV^{-3}$, and $F_1=F_2=-0.0263 GeV^6$. Here it is to be noted that $F_{1,2}$ for 2P are negative, which is physically expected from the nodal structure, but they enter as $F_{1}^2$ and $F_{2}^2$ in $|\bar{M}_{fi}|^2$, and contribute positively. Thus the values, $|N_AF_1|$=0.2036 for 1P, and $|N_A F_1|$=0.3362 for 2P, give a factor of 1.65 in the overlap , which leads to a factor $\sim 2.7$ in the cross section.

\begin{figure}[ht!]
 \centering
 \includegraphics[width=10cm,height=6cm]{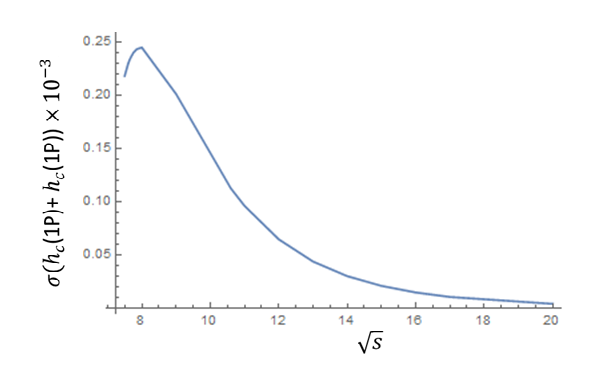}
 \caption{Plot of cross section for $e^- e^+\rightarrow \gamma^*\gamma^*\rightarrow h_c(1P)+h_c(1P)$ (in fb) versus $\sqrt{s}$ (in GeV)}
 \end{figure}
 
The plot of the cross section for $e^- e^+\to h_c(1P)+h_c(1P)$ versus $\sqrt{s}$ in Fig.3 displays a peak near the threshold followed by the asymptotic fall at higher energies.

\subsection{Sensitivity of predicted cross sections}

To estimate the sensitivity of our cross sections predictions for $\chi_{c1}+\eta_c$, and $h_c+h_c$ to model input parameters, we varied the charm-quark mass $m$, the QCD scale parameter $\Lambda$, and the flavour-independent spring constant $\omega_0$ around their central values fixed from charmonium spectroscopy\cite{eshete19,vaishali21a}. The results are summarized in Tables 1 and  4. The variations of quark mass $m$ and the QCD scale parameter $\Lambda$ lead to relatively modest changes in the predicted cross sections, typically at the level of a few to about ten percent. This indicates that the results for both the channels are reasonably stable against changes in the fundamental QCD inputs entering the model.  

A stronger sensitivity is observed with respect to the spring constant $\omega_0$, particularly for the $h_c+h_c$ channel. This behaviour is expected since $\omega_0$ (along with quark mass $m$ and QCD length scale $\Lambda$) directly control the spatial scale of the bound-state wavefunctions through the inverse-range parameter $\beta$. And the effect is more pronounced in the double P-wave channel, where higher powers of the internal momentum enter the production amplitude. The strong sensitivity of the $h_c+h_c$ channel to $\omega_0$ can be understood analytically by analyzing the scaling behaviour of $h_c+h_c$ cross section with $\omega_0$. The 3D integrals and the BS normalizers entering into Eq.(28) for $|\bar{M}_{fi}|^2$, scale with the inverse range parameter $\beta$ as $F_1=F_2\sim \beta^6$, and $N_A=N_{A'}\sim \beta^{-5/2}$. And since the inverse range parameter in the 3D radial BS wave functions, $\beta\sim \omega_{0}^{1/2}$, it leads to the scaling behaviour, $\sigma(h_c+h_c)\sim \omega_{0}^7$, arising from the momentum weighted double P-wave Bethe-Salpeter vertices. Thus the variation in $\omega_0$ significantly modifies the absolute magnitude of the $h_c+h_c$ cross section, while the qualitative energy dependence and the relative hierarchy of the two channels remain largely unaffected. This enhanced sensitivity of the $h_c+h_c$ channel to $\omega_0$ reflects the intrinsic dependence of double P-wave production on the momentum-space structure of the bound-state wavefunctions\cite{eshete19,vaishali21a}. In this sense, the $h_c+h_c$ process may serve as a sensitive probe of the confinement dynamics encoded in the Bethe–Salpeter kernel. However for the $\chi_{c1}+\eta_c$ channel, the cross section scales as $\sigma(\chi_{c1}+\eta_c)\sim \omega_{0}^3$, due to the fact that $N_A G_1\sim \beta_{A}^{3/2}$, while $N_P G_2\sim \beta_{P}^{1/2}$, due to which $\sigma \sim (N_A G_1)^2 (N_P G_2)^2\sim \beta_{A}^3 \beta_P \sim \omega_{0}^2 \omega_{0}\sim \omega_{0}^3$, leading to much smaller sensitivity.

\section{Discussions}
In this work, we have investigated the exclusive double charmonium production processes:
$e^- e^+ \to \gamma^\ast \gamma^\ast \to \chi_{c1} + \eta_c$ and
$e^- e^+ \to \gamma^\ast \gamma^\ast \to h_c + h_c$
at the center-of-mass energy $\sqrt{s}=10.6~\mathrm{GeV}$ within relativistic $4\times4$ Bethe--Salpeter equation framework under the Covariant Instantaneous Ansatz,
which provides a consistent and covariant treatment of relativistic bound-states. Both channels studied here proceed purely through quark-rearrangement diagrams, since photon-fragmentation contributions are forbidden by charge-conjugation and spin-parity constraints. We obtain cross sections of order $10^{-3}$fb with $\sigma(\chi_{c1}(1P)+\eta_c(1S))= 1.315\times 10^{-3}$fb, and $\sigma(h_c(1P)+h_c(1P))=0.113\times 10^{-3}$fb. The quoted uncertainties represent only the sensitivity of the predictions to variations of the fitted model parameters. They do not include systematic uncertainties associated with the approximations of the present BSE-CIA framework. This strong suppression of these cross sections indicates the absence of fragmentation enhancement and is a pointer to the role of relativistic quark-rearrangement dynamics that is encoded in the Bethe–Salpeter wave functions. Our results for $\chi_{c1}+\eta_{c}$ channel are consistent with NRQCD expectations at the qualitative level, although no experimental data are currently available for these channels. However for $h_c+h_c$ channel, Table 5 shows BSE-CIA predicts $\sigma(1P+1P)=0.113^{+0.028}_{-0.023}\times 10^{-3}$fb against NRQCD value: $(0.19\pm 0.02)\times 10^{-3}$ fb. The suppression of BSE value may be due to the reason that the BSE framework naturally suppresses P-wave production relative to NRQCD due to the relativistic vertex structure in BSE framework, that reduce the overlap integral relative to the NRQCD non-relativistic approximation.

\bigskip

To test the strength of our predictions, we examined the sensitivity of the cross sections to moderate variations of key model parameters, namely the charm-quark mass m, 
the QCD scale parameter $\Lambda$ (entering the running coupling $\alpha_S$), and the flavour-independent spring constant $\omega_0$ that enter into the confining kernel. We find that variations of fundamental QCD parameters, $m$ and $\Lambda$ within commonly accepted ranges lead to only modest changes in the predicted cross sections, typically at the level of a few to about ten percent, while a stronger dependence is observed with respect to variations of $\omega_0$. This effect is particularly visible in the $h_c+h_c$ channel, where double P-wave production introduces higher powers of internal momentum in the amplitude, leading to the scaling behaviour of its cross section, $\sigma(h_c+h_c)\sim \omega_{0}^7$, reflecting the greater sensitivity of the double P-wave process to the confinement parameter $\omega_0$ as demonstrated in the variance analysis.

The BSE-CIA further provides the first prediction for $\sigma(h_c(2P)h_c(2P)) = 0.298 ^{+0.075}_{-0.063} \times 10^{-3}$fb, which is 2.6 times larger than the 1P+1P result. This enhancement can be understood in terms of the dynamical quantities that govern P-wave production in the BSE framework. This contrasts with the behavior seen in the $\chi_{c1}+\eta_c$
channel, where $\sigma(1P+1S) > \sigma(2P+2S)$. The distinction arises from the momentum-space structure of the respective overlap integrals. Here we would like to mention that in NRQCD, the production rate of P-wave quarkonia is controlled by $|R'_{nP}(0)|^2$, which measures the strength of the P-wave bound-state wave function at short distances. However, in the present BSE formalism, the analogous role is played by the momentum-weighted radial overlap integral $F_1=\int \frac{d^3\hat{q}}{(2\pi)^3}\hat{q}^2\phi_A(\hat{q})$ together with the normalization factor, $N_A$. In the $h_c+h_c$
amplitude, the production integral $F_1$ carries an explicit $\hat{q}^2$  weight, arising from the $(\hat{q}.\epsilon)$ factors in the $h_c$ BS vertex function, Eq.(22).  This momentum weighting preferentially samples the high $|\hat{q}|$ region of the wave function, where the 2P state has relatively greater support beyond its radial node. As a result, the effective production quantity $|N_A F_1|$  is numerically larger for $h_c(2P) (12.775\times 0.0263 = 0.336)$, than for $h_c(1P) (13.664\times 0.0149 = 0.204)$ leading to $\sigma(2P+2P) > \sigma(1P+1P)$. However, the qualitative energy dependence and overall hierarchy of the predicted channels remain stable under reasonable parameter variations, indicating that the principal physical conclusions of the present work are robust. Since there are no independent theoretical calculations or experimental measurements that presently exist for the $h_c(2P)+h_c(2P)$ channel, this prediction awaits future verification.
 
In contrast, the $\chi_{c1}+\eta_c$ amplitude involves the unweighted integrals $G_1 = \int \frac{d^3\hat{q}}{(2\pi)^3}\phi_A(\hat{q})$, and $G_2 = \int \frac{d^3\hat{q}'}{(2\pi)^3}\phi_P(\hat{q}')$, which contain no extra momentum-weighting factor. Without this weight factor, the nodal cancellation in the 2P and 2S wave functions is not compensated by any high $\hat{q}^2$ enhancement, and the integrals are numerically smaller for excited states than for ground states. This is confirmed by the values: $|N_A G_1| = 0.0409$
 for $\chi_{c1}(1P)$ versus 0.0383 for $\chi_{c1}(2P)$, and $|N_P G_2|$ = 0.0527 for $\eta_c(1S)$ versus 0.043 for $\eta_c(2S)$. Both factors together give $\sigma(1P+1S) > \sigma(2P+2S)$
 Thus the contrasting hierarchies in the two channels are a direct consequence of the different momentum structures of their respective BS vertex functions, reflecting the deeper relativistic spin dynamics encoded in the BSE-CIA framework.

The energy dependence for the two channels in Figs. 2 and 3 shows a similar qualitative behaviour. The $\chi_{c1}+\eta_c$  and $h_c+h_c$ cross sections decreases smoothly and monotonically with $\sqrt{s}$ approaching a power-law falloff which is a few GeV above threshold. The cross sections reveal the asymptotic scaling  ($\sim 1/s^2$) leading to a falloff in the cross section. This is in contrast to \cite{liao24,liao22}, where the cross section for production of double P-wave charmonia shows a non-monotonic lower-energy region followed by an asymptotic decrease.

An important outcome of our study is that, within a unified BSE framework whose parameters are fixed entirely from spectroscopy\cite{eshete19,vaishali21a}, one can obtain self-consistent predictions for highly suppressed exclusive production channels. The predicted cross sections are small ($\approx 10^{-3}$ fb), but their internal hierarchy and energy dependence follow physically transparent patterns, that are largely a result of radial wave-function structure (see \cite{eshete19,vaishali21a} for details) and relativistic vertex dynamics. Our results highlight the role of relativistic bound-state structure in exclusive heavy-quarkonium production and provide quantitative benchmarks for future high-luminosity measurements.

The qualitative agreement with NRQCD hierarchies indicates that certain features of double-charmonium production are model-independent consequences of heavy-quark dynamics, while quantitative differences highlight the role of relativistic bound-state effects beyond the leading non-relativistic approximations. In this sense, the two approaches should be viewed as complementary: NRQCD organizes short-distance dynamics systematically, whereas the BSE framework encodes full relativistic wave-function structure and its impact on amplitudes. 

From a phenomenological point of view, although the predicted rates are below current experimental sensitivity, the rapid luminosity increase at Belle II and future $e^-+e^+$
facilities may allow meaningful upper limits to be set. More importantly, the present analysis demonstrates that suppressed two-photon channels can serve as precision probes of quarkonium structure. Extending this program to higher energies and next-to-leading-order corrections would provide a systematic pathway toward a deeper understanding of relativistic effects in heavy-quark bound states.

\end{document}